\documentclass[aps,prl,twocolumn,superscriptaddress,showpacs]{revtex4-1}

\usepackage{graphicx}
\usepackage{amsmath}
\usepackage{cleveref}
\usepackage{siunitx}
\usepackage{color} 
\usepackage{setspace}

\usepackage{amssymb}
\usepackage{booktabs}

\usepackage[bookmarks=true,colorlinks=true,urlcolor=blue,linkcolor=blue,citecolor=blue,breaklinks]{hyperref}

\begin{document}

\sloppy

\title{Ultrafast Transient Dynamics of Adsorbates on Surfaces Deciphered:\\
The Case of CO on Cu(100)}








\newcommand*{\IFS}[0]{{
Center of Excellence for Advanced Materials and Sensing Devices, Institute of Physics, Bijeni\v{c}ka 46,
10000 Zagreb, Croatia}}

\newcommand*{\DIPC}[0]{{
Donostia International Physics Center (DIPC),
Paseo Manuel de Lardizabal 4, 20018 Donostia-San Sebasti\'an, Spain}}

\newcommand*{\FUBER}{{
Institut f{\"u}r Chemie und Biochemie, Freie Universit{\"a}t Berlin,
Takustrasse~3, 14195 Berlin, Germany}}

\newcommand{\CFM}[0]{{
Centro de F\'{\i}sica de Materiales (CSIC-UPV/EHU), Paseo Manuel de Lardizabal 5, 20018 Donostia-San Sebasti\'an, Spain}}

\newcommand{\QUIMICAS}[0]{{
Departamento de F\'{\i}sica de Materiales, Facultad de Qu\'{\i}micas UPV/EHU,
Apartado 1072, 20080 Donostia-San Sebasti\'an, Spain}}

\author{D. Novko}
\email{dino.novko@gmail.com}
\affiliation{\IFS}
\affiliation{\DIPC}
\author{J. C. Tremblay}
\affiliation{\FUBER}
\author{M. Alducin}
\affiliation{\CFM}
\affiliation{\DIPC}
\author{J. I. Juaristi}
\affiliation{\QUIMICAS}
\affiliation{\CFM}
\affiliation{\DIPC}

\begin{abstract}

Time-resolved vibrational spectroscopy constitutes an invaluable experimental tool for monitoring hot-carrier-induced surface reactions. However, the absence of a full understanding of the precise microscopic mechanisms causing the transient spectral changes has limited its applicability. Here we introduce a robust first-principles theoretical framework that successfully explains both the nonthermal frequency and linewidth changes of the CO internal stretch mode on Cu(100) induced by femtosecond laser pulses. Two distinct processes engender the changes:  electron-hole pair excitations underlie the nonthermal frequency shifts, while electron-mediated vibrational mode coupling gives rise to linewidth changes. Furthermore, the origin and precise sequence of coupling events are finally identified. 


\end{abstract}

\maketitle

%
%

One of the ultimate goals in surface science is to comprehend the fundamental processes that bring about the specific timescales of surface reactions\,\cite{bib:petek00,bib:arnolds10,bib:dellangela13}. To acquire such a time-resolved insight, numerous experiments have studied the ultrafast elementary motions of adsorbates on metal surfaces by means of time-dependent techniques, including vibrational motion\,\cite{bib:morin92,bib:germer93,bib:watanabe10}, molecular desorption\,\cite{bib:busch95,bib:struck96,bib:bonn99,bib:bonn00,bib:inoue16}, diffusion\,\cite{bib:stepan05,bib:backus05}, and dissociation\,\cite{bib:lane06}. In these experiments, the transient condition achieved by the use of intense femtosecond laser pulses initiates the energy exchange mechanisms between the laser-excited surface electrons and the vibrational modes of the adsorbates and surface lattice. 

In time-resolved infrared (IR) spectroscopy experiments, the initial adsorbate dynamics commenced by the pump pulse is directly probed by tracking the frequency shift and linewidth changes of the IR-active internal stretch (IS) mode. In all adsorbate-surface systems investigated thus far [e.g., CO/Ru(001)\,\cite{bib:bonn00}, NO/Ir(111)\,\cite{bib:lane06}, CO/Pt(111)\,\cite{bib:fournier04,bib:watanabe10}, CO/Cu(100)\,\cite{bib:inoue16}], the IS frequency mode exhibits an initial redshift followed by a rapid blueshift. 
However, the origin of such a characteristic behavior is still not fully understood. In the early works it was ascribed to anharmonic coupling with other low-energy (LE) modes\,\cite{bib:persson85,bib:persson86,bib:bonn00}, while later energy transfer from the laser-excited hot electrons to the adsorbate motion via nonadiabatic coupling (NC) was also considered\,\cite{bib:lane06,bib:ueba08b,bib:watanabe10,bib:inoue16}. Nonetheless, the lack of a general quantitative theory  effectively prevents us from harnessing the full potential of time-resolved vibrational spectroscopy, as well as from extracting the information about subpicosecond dynamics of surface reactions buried within.

In this Letter, we introduce a general first-principles theoretical framework that allows us to calculate directly the vibrational spectra changes and identify the specific mechanisms behind them. Our theory relies on a recently developed approach based on many-body and density functional perturbation theories\,\cite{bib:novko18} that we extend here to treat nonequilibrium conditions. Our first-principles formalism accounts for electron-phonon coupling processes up to second order, including electron-hole pair excitations (the first-order NC) as well as vibrational intermode coupling due to the indirect interaction with hot electrons. This so-called electron-mediated phonon-phonon coupling (EMPPC) allows the interaction of phonons with incommensurate frequencies, even in the absence of direct anharmonic coupling\,\cite{bib:novko18}.

\begin{figure}[!t]
\includegraphics[width=0.49\textwidth]{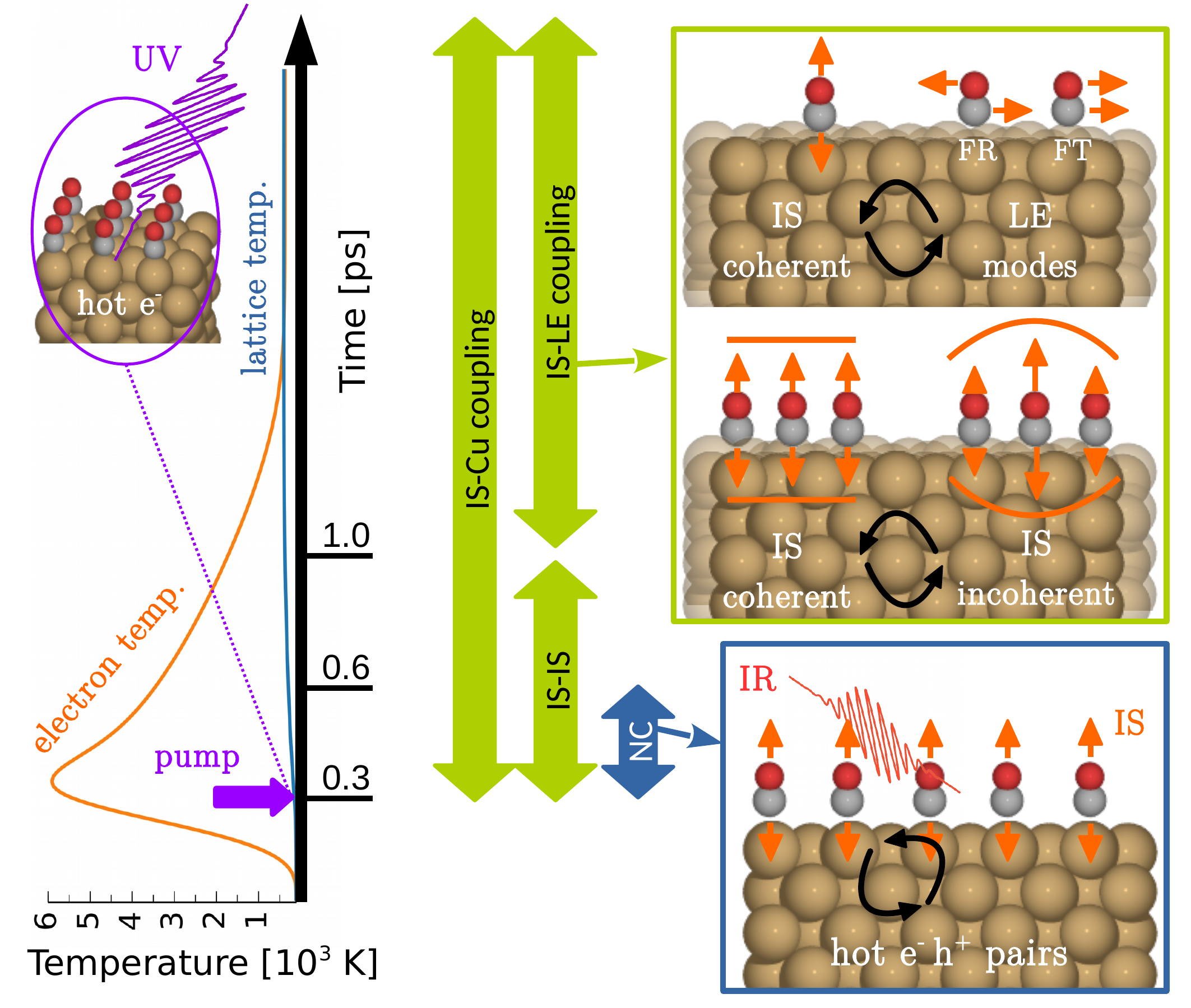}
\caption{\label{fig:fig1}Schematic chronograph of the coupling mechanisms underlying the IS frequency shifts of CO on Cu(100). (Left) Example of the time evolution followed by the electron and phonon temperatures when an UV pump laser starting at $t=0.3$\,ps heats the surface. (Right) Time evolution of the hot-carrier-induced CO dynamics. First-order NC process is active between around 0.3 and 0.6\,ps (blue frame and arrow). Electron-mediated vibrational mode couplings start at around 0.3\,ps and operates until equilibration (green frame). The specific time intervals are indicated by corresponding green arrows.}
\end{figure}

Here we apply this theoretical framework to investigate the 
early stage dynamics of nonthermal CO adsorbates (i.e., a molecular overlayer) on Cu(100). This system, which is considered as the benchmark system for electron-induced dynamics at surfaces, has recently been monitored with time-resolved sum-frequency generation spectroscopy with unprecedented subpicosecond resolution\, \cite{bib:inoue16}.
However, it is our quantitative and predictive theoretical method that finally unveils the microscopic processes behind the reported nonthermal frequency shifts and the accompanying linewidth changes, while it also establishes the specific sequence and strength of the nonadiabatic and intermode coupling mechanisms involved. 
More specifically, we show that the coupling mechanisms behind the transient frequency and linewidth changes are in fact different. The first-order NC contribution describing nonthermal charge transfer dominates the frequency shifts and acts 
rapidly.
In contrast, linewidth changes are mostly due to EMPPC, which operates on longer timescales, up to tens of picoseconds.
Furthermore, the explicit sequence of events affecting the transient adsorbate motion are unambiguously identified and quantified from our calculations, as illustrated in Fig.\,\ref{fig:fig1}. On the subpicosecond timescale, the coherent IS phonon mode (probed by the IR light) is strongly coupled to the activated incoherent IS phonon modes. 
Here, coherent and incoherent refer to the in phase ($\mathbf{q}\approx 0$) IS phonon mode and to the incoherent averaging of all IS adsorbate phonon modes with finite momentum (i.e., $\mathbf{q}>0$), respectively.
After around 1\,ps, this coupling weakens and the coherent IS mode couples predominantly to the adsorbate LE modes. Interestingly, energy exchange between the IS mode and surface atom motions is important at all times, due to the sheer number of these modes. 
Our results challenge the current belief that only intermode coupling to the LE adsorbate modes are involved in the subpicosecond dynamics upon femtosecond laser excitation\,\cite{bib:persson85,bib:persson86,bib:bonn00,bib:lane06,bib:arnolds10,bib:inoue16}, while the importance of the surface and incoherent IS motions is generally understated.

\begin{figure}[!t]
\includegraphics[width=0.49\textwidth]{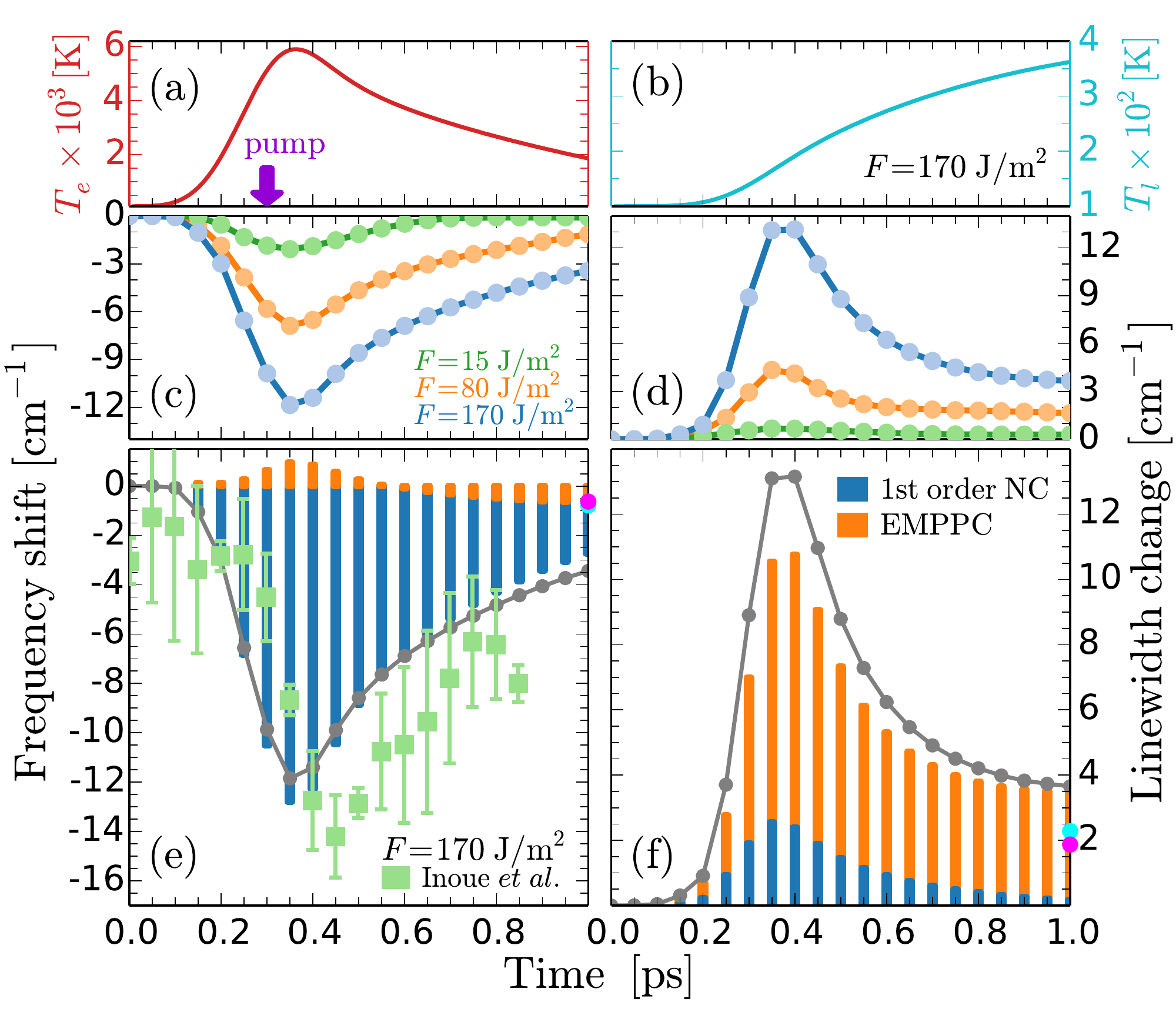}
\caption{\label{fig:fig2}Transient changes in the CO/Cu(100) system induced by 400\,nm pump-laser pulses with 150\,fs duration. (a) Electron $T_e(t)$ and (b) lattice $T_l(t)$ temperatures as a function of time for an absorbed fluence $F=170$\,J/m$^2$. Initial temperature is 100\,K. Time-dependent (c) frequency and (d) linewidth changes of the IS mode under nonthermal condition induced by the UV pump at $t=0.3$\,ps. Blue, orange, and green represent the results for different $F$. (e),\,(f) Contribution-resolved analysis of frequency and linewidth changes when $F=170$\,J/m$^2$. Blue represents the contribution coming from the first-order NC term and orange shows the EMPPC contribution, while gray circles are the sums of the two. Light blue and magenta points show the corresponding results for $t=50$ and 100\,ps, respectively. Green squares are experimental frequency shifts obtained from Refs.\,\cite{bib:inoue16}.} 
\end{figure}

Within many-body perturbation theory, the phonon frequency renormalization and linewidth due to NC are given in terms of the real and imaginary parts of the phonon self-energy $\pi_{\nu}(\omega;t)$, i.e., $\omega^2-\omega_{\nu}^2=2\omega_{\nu}\mathrm{Re}\,\pi_{\nu}(\omega;t)$ and $\gamma_{\nu}=-2\mathrm{Im}\,\pi_{\nu}(\omega;t)$, respectively ($\nu$ denotes the IS phonon mode)\,\cite{bib:novko16a,bib:giustino17,bib:caruso17,bib:ferrante18} (see Supplemental Material\,\cite{bib:si} for computational details). 
The real part encodes how the C-O interaction and concomitant IS frequency adapt to the environment brought out of equilibrium by the external light source. The imaginary part reveals how the IS mode decays because of inelastic collisions with the hot electrons and phonons; therefore, it can be directly related to the IS mode transient lifetime $\tau_{\nu}=\hbar/\gamma_{\nu}$ or to the time-resolved full width at half maximum of the CO IS mode spectral resonance, which are often reported in the experiments\,\cite{bib:bonn00,bib:lane06,bib:watanabe10}.
In the present Letter, we consider that the phonon self-energy consists of first- and second-order terms in the electron-phonon coupling, i.e., $\pi_{\nu}(\omega;t)=\pi^{[1]}_{\nu}(\omega;t)+\pi^{[2]}_{\nu}(\omega;t)$\,\cite{bib:novko16a,bib:novko18}. 
These are simultaneously the dominant interband  and intraband
contributions, respectively.
The former term accounts for electron-hole pair excitation processes\,\cite{bib:novko16a} including nonthermal charge transfer to unoccupied states (e.g., antibonding CO states). The latter term is responsible for the EMPPC\,\cite{bib:novko18}. These two features are known to be instrumental for explaining the long wavelength part of the phonon spectrum in metallic bulk systems\,\cite{bib:marsiglio92,bib:saitta08,bib:giustino17,bib:ferrante18}. The processes behind $\pi_{\nu}^{[1]}$ depend on the electron distribution, while $\pi_{\nu}^{[2]}$ depends on both electron and phonon distributions\,\cite{bib:novko18}. In our theory, the first- and second-order contributions depend indirectly on time via the electron $T_e(t)$ and lattice $T_l(t)$ temperatures describing the excited electron and phonon distributions. These are brought out of equilibrium by coupling with the pump-laser pulse, and their time evolution is obtained from the two temperature model (TTM)\, \cite{bib:allen87,bib:lin08}. Thus far, the TTM was mostly used to complement molecular dynamics simulation of 
isolated adsorbates on surfaces and, therefore, only surface motion was included in $T_l(t)$\,\cite{bib:springer94,bib:saalfrank06}. In our formulation, however, the EMPPC term $\pi_{\nu}^{[2]}$ contains information on intermode couplings between the IR-excited IS mode and all other modes in the system. These include the incoherent IS motion, the external stretch (ES) mode, and frustrated rotations and translations (FR and FT), as well as surface motion\,\cite{bib:novko18}. Consequently, $T_l(t)$
represents the total lattice temperature coming from all excited modes in CO/Cu(100) (see Supplemental Material\,\cite{bib:si}). As an example, Figs.\,\ref{fig:fig2}(a) and \ref{fig:fig2}(b) show the time evolution of $T_e(t)$ and $T_l(t)$ when the experimental 400\,nm pump laser, with 170\,J/m$^2$ absorbed fluence and 150\,fs duration, heats the surface electrons at 0.3\,ps. Altogether, this combination of first-principles many-body perturbation theory up to second-order with the TTM, which accounts for the time-dependent excited electron and phonon distributions\,\footnote{We stress that the effective temperature models (such as the TTM) are unable to capture the nascent nonequilibrium electron distribution below the timescale of the electron-electron scattering time $\tau_{ee}$\,\cite{bib:sentef13}. Nevertheless, the time resolution of the experiment studied here is well above the typical formation time of the hot Fermi-Dirac distribution with effective temperatures present in the TTM (i.e., pulse duration  $> \tau_{ee}$).}, constitutes a substantial improvement upon state-of-the-art theories. These include \textit{ab initio} first-order vibrational damping rate theories\,\cite{bib:forsblom07,bib:askerka16,bib:rittmeyer17} and parametric models of pure dephasing mediated by anharmonic coupling\,\cite{bib:persson85,bib:persson86} or electron-hole pair creation\,\cite{bib:morawitz87,bib:persson02,bib:ueba08b}.

Figures\,\ref{fig:fig2}(c) and \ref{fig:fig2}(d) show, respectively, 
the frequency and linewidth changes of the IS mode, i.e., $\delta\omega_{\mathrm{IS}}$ and $\delta\gamma_{\mathrm{IS}}$ (see Supplemental Material\,\cite{bib:si} for more details), calculated at various absorbed laser fluences $F$ of the experimental pump-laser pulse.
The frequencies are characterized by a redshift over the course of the first 0.3\,ps, followed by a blueshift. The phonon linewidths increase at short times, reaching their highest values when the frequency redshifts are at the maximum, and decrease at longer times. Also, both $\delta\omega_{\mathrm{IS}}$ and $\delta\gamma_{\mathrm{IS}}$ are more pronounced as the absorbed fluence becomes more intense. A good example of the remarkable agreement between the calculated $\delta\omega_{\mathrm{IS}}$ and the experimental shifts is shown in Fig.\,\ref{fig:fig2}(e) for the specific absorbed fluence of 170\,J/m$^2$\,\cite{bib:inoue16}.
Importantly, the time dependence of both $\delta\omega_{\mathrm{IS}}$ and $\delta\gamma_{\mathrm{IS}}$
obtained here shows a very close resemblance to results of ultrafast vibrational spectroscopy experiments of other adsorbate-surface systems, e.g., CO/Ru(001)\,\cite{bib:bonn00}, NO/Ir(111)\,\cite{bib:lane06}, or CO/Pt(111)\,\cite{bib:fournier04,bib:watanabe10}. Since these nonthermal effects are universal and represent the fingerprint of ultrafast surface vibrational dynamics\,\cite{bib:arnolds10}, the applicability of our theory can be extended to a variety of adsorbate-surface systems and thus help to elucidate other intriguing surface reactions as well. 

In Figs.\,\ref{fig:fig2}(e) and \ref{fig:fig2}(f), the contributions
to $\delta\omega_{\mathrm{IS}}$ and $\delta\gamma_{\mathrm{IS}}$ arising from the first-order NC processes (i.e., $\pi^{[1]}$) and from the EMPPC (i.e., $\pi^{[2]}$) are depicted as blue and orange bars,
respectively.
Surprisingly, the mechanisms ruling the frequency shift and the lifetime are different. Thus, 
$\delta\omega_{\mathrm{IS}}$ is dominated during the first picosecond by the electron-hole pair excitations,
while the EMPPC brings only a small contribution. Conversely, $\delta\gamma_{\mathrm{IS}}$ is dominated by the EMPPC mechanism
from the outset. At later times, i.e., when $t=50$ and 100\,ps, the EMPPC is the only contributor to both $\delta\omega_{\mathrm{IS}}$ and $\delta\gamma_{\mathrm{IS}}$ (see also Table\,\ref{tab:table1}). These findings are in contradiction with the usual assumption that both transient frequency and linewidth changes of the IS mode can be explained by means of a single dominating nonthermal mechanism. In particular,
the experimental observations are often attributed either to pure dephasing induced by anharmonic coupling between the IS and a single LE mode\,\cite{bib:persson85,bib:persson86,bib:bonn00}, or to nonthermal charge transfer from metal occupied states to an antibonding state of the CO adsorbate\,\cite{bib:lane06}.
In other words, while our result confirms the view that $\delta\omega_{\mathrm{IS}}$ is mainly triggered by the transient population of the antibonding state\,\cite{bib:lane06}, it challenges the belief that both $\delta\omega_{\mathrm{IS}}$ and $\delta\gamma_{\mathrm{IS}}$ are underlain by the same process.

\begin{table}[!b]
\caption{\label{tab:table1}Contribution- and mode-resolved analyses of the IS mode nonthermal frequency $\delta\omega_{\mathrm{IS}}$ and linewidth $\delta\gamma_{\mathrm{IS}}$ changes are shown for several
snapshots when $F=170$\,J/m$^2$. The upper part of the table resolves the contributions associated with the first-order NC term $\pi^{[1]}$ (electron-hole pair excitations) and the EMPPC term $\pi^{[2]}$.
In the bottom part of the table, the EMPPC term is further separated in its mode-resolved contributions, including intermode coupling between the coherent (in phase) IS mode and
the incoherent IS, ES, FR, FT, and Cu modes of the system.
All reported frequency shifts and linewidth changes are in cm$^{-1}$.}
\begin{ruledtabular}
\begin{tabular}{ccccccc}
 & \multicolumn{2}{c}{$t=0.4$\,ps} & \multicolumn{2}{c}{$t=1$\,ps} & \multicolumn{2}{c}{$t=50$\,ps} \\
 \cline{2-3}\cline{4-5}\cline{6-7}
 & $\delta\omega_{\mathrm{IS}}$ & $\delta\gamma_{\mathrm{IS}}$ & $\delta\omega_{\mathrm{IS}}$ & $\delta\gamma_{\mathrm{IS}}$ & $\delta\omega_{\mathrm{IS}}$ & $\delta\gamma_{\mathrm{IS}}$ \\
\hline
\rule{0pt}{3ex}
$\pi^{[1]}$    & -12.27 &  2.40 & -2.75 & 0.18 & -0.16 & -0.01 \\
$\pi^{[2]}$    &   0.88 & 10.76 & -0.69 & 3.48 & -0.62 &  2.30 \\
\hline
\rule{0pt}{3ex}
IS             &   0.19 &  3.03 &  0.02 & 0.20 &  0.00 & 0.03 \\
ES             &   0.06 &  0.22 &  0.01 & 0.04 &  0.00 & 0.02 \\
FR             &   0.03 &  0.91 & -0.13 & 0.68 & -0.15 & 0.39 \\
FT             &   0.00 &  0.75 & -0.14 & 0.72 & -0.15 & 0.44 \\
Cu          &   0.60 &  5.85 & -0.45 & 1.84 & -0.32 & 1.42 \\
\end{tabular}
\end{ruledtabular}
\end{table}

Furthermore, since $\pi^{[1]}$ depends on the electron distribution excited by the laser through $T_e$ and $\pi^{[2]}$ on both the distributions of excited electrons and phonons via $T_e$ and $T_l$, our theory 
delivers a direct correlation between nonthermal changes and the time-dependent temperatures.
In particular, the first-order NC mechanism contributes to a softening of the IS bond as $T_e$ increases, which causes the observed initial redshift in the IS frequency that decreases in magnitude as $T_e$ starts to decrease. Additionally, the increase in the linewidth due to the first-order NC also follows the changes in $T_e$.
In contrast, changes due to the EMPPC are influenced by both temperature effects. The latter manifests in the time dependence of $\delta\gamma_{\mathrm{IS}}$ as an increase up to around 0.4\,ps, which correlates with $T_e(t)$, followed by a decrease to a finite offset with respect to the original value, which reflects the increase experienced by the lattice temperature in this interval. Stated differently, both the excited electrons and phonons, which are described through $T_e$ and $T_l$ in the many-body response term $\pi^{[2]}$, trigger a similar, positive trend in $\delta\gamma_{\mathrm{IS}}$. 
Interestingly, these two temperature effects compete in the EMPPC contribution to $\delta\omega_{\mathrm{IS}}$: an increase of $T_e$ induces a blueshift, while an increase of $T_l$ induces a redshift (see Supplemental Material\,\cite{bib:si} for more details).

Our theory further allows us to resolve the different mode contributions in $\delta\omega_{\mathrm{IS}}$ and $\delta\gamma_{\mathrm{IS}}$ (Table\,\ref{tab:table1}) at 
different instants during the thermal equilibration process. Since $\delta\omega_{\mathrm{IS}}$ is mostly dominated by the first-order NC term during the time of interest spanned by the experiment, we focus our discussion on the relevant time-resolved intermode couplings responsible for $\delta\gamma_{\mathrm{IS}}$. We select three instants representative of the different thermal conditions reached during the equilibration stage: $t=0.4$\,ps ($T_e\gg T_l\approx 100\,$K), $t=1$\,ps ($T_e > T_l>100\,$K), and $t=50$\,ps ($T_e\approx T_l$).
At the onset of nonthermal vibrational dynamics, the EMPPC mechanism is mostly prompted by the nonequilibrium electron distribution, since $T_e\gg T_l$. During that period, the prevailing mode coupling is between the probed (coherent) IS mode and the incoherent IS modes.
We dub this mechanism ``intramode coupling'', and it can be regarded as the dephasing process of the IS mode. This unexpected result is due to a very rapid increase of the electron-mediated IS-IS coupling strength with the rise of $T_e$, when compared to other molecular intermode coupling strengths (cf. Supplemental Material\,\cite{bib:si}). In other words, when $T_e\gg T_l$, the small energy gap of around 10\,meV that exists between coherent and incoherent IS modes\,\cite{bib:novko18} 
is efficiently compensated by the broad distribution of hyperthermal electrons, leading to a strong electron-mediated IS-IS scattering. When $T_e>T_l>100\,$K ($t\approx1$\,ps), but also when the system is almost thermalized ($t\gg 1$\,ps), the time-dependent EMPPC processes arise prevalently from the excited LE modes. In particular, the FR and FT modes are found to be the major contributors to the nonthermal changes. Aside from the molecular modes, the remaining Cu surface motions are also crucial for describing $\delta\gamma_{\mathrm{IS}}$.
Once the electron-mediated IS-Cu mode coupling mechanism has been activated, it remains active until the final equilibration. This originates from the abundance of surface Cu modes at low energy that are easily excited by heat.

This Letter explains thus far unresolved intricacies of nonadiabatic and intermode couplings at surfaces under nonequilibrium conditions. Contrary to common understanding, the CO internal stretch mode undergoes different mode coupling mechanisms on the subpicosecond and the picosecond timescales.
Namely, the relaxation is first driven by the dephasing process consisting of the IS-IS coupling ($t<1$\,ps), after which the coupling with the lateral modes prevails.
Equally intriguing, transient frequency shifts turn out to be underlain by electron-hole pair excitations, while linewidth changes are mostly governed by the electron-mediated phonon-phonon coupling. 
This implies that the C-O interaction, which is in the end responsible for the IS frequency, is more sensitive to the transient excitations created in the electronic system, while it is the coupling to the excited phonon modes that contributes more to the IS linewidth. All in all, the presented theory is destined to be the theoretical counterpart in future vibrational spectroscopy investigations. The time-resolved nanoscopic insights that this theory can provide are not only fundamental to the development of  vibrational spectroscopy at surfaces, but additionally to advance in our goal of controlling surface reactions at the molecular level.

\vspace{1mm}

We thank B. Gumhalter, P. Saalfrank, and R. D\'iez Mui\~no for useful discussions and comments. D.\,N. acknowledges financial support from the European Regional Development Fund for the ``Center of Excellence for Advanced Materials and Sensing Devices'' (Grant No. KK.01.1.1.01.0001) as well as from Donostia International Physics Center (DIPC).
J. C. T. is grateful to the Deutsche Forschungsmeinschaft for funding through grant TR1109/2-1.
M.\,A. and J.\,I.\,J. acknowledge the Spanish Ministerio de Econom\'{\i}a, Industria y Competitividad Grant No. FIS2016-76471-P. Computational resources were provided by the DIPC computing center.

\bibliography{emppct}

\end{document}